\documentclass[aps,prl,twocolumn,superscriptaddress,preprintnumbers,floatfix]{revtex4}
\usepackage{graphics,colordvi}
\usepackage{amsmath,amssymb,epsfig}

\makeatletter
\newcommand{\fmslash}[2][0mu]{%
  \mathchoice
    {\fmsl@sh\displaystyle{#1}{#2}}%
    {\fmsl@sh\textstyle{#1}{#2}}%
    {\fmsl@sh\scriptstyle{#1}{#2}}%
    {\fmsl@sh\scriptscriptstyle{#1}{#2}}}
\newcommand{\fmsl@sh}[3]{%
  \m@th\ooalign{$\hfil#1\mkern#2/\hfil$\crcr$#1#3$}}

\newcommand{\tr}{\hbox{tr}}

\def\eq#1\en{\begin{equation} #1 \end{equation}}

\def\eqa#1\ena{\begin{eqnarray} #1 \end{eqnarray}}

\newcommand{\GeV}{\mbox{\rm GeV}}

\usepackage{graphicx,color}

\begin{document}

\title[High energy cosmic rays experiments inspired by noncommutative quantum field theory]{High energy cosmic rays experiments inspired by noncommutative quantum field theory}
\author{Josip Trampeti\'{c}}
\affiliation{Rudjer Bo\v skovi\' c Institute, Bijeni\v cka 54, HR-10.002 Zagreb, P.O.Box 180, Croatia}
\email{josipt@rex.irb.hr}
\date{\today}


\begin{abstract}
Phenomenological analysis of the covariant $\theta$-exact noncommutative (NC)
gauge field theory (GFT), inspired by high energy cosmic rays experiments, is performed in the framework of the inelastic neutrino-nucleon scatterings, plasmon and $Z$-boson decays into neutrino pair, the Big Bang Nucleosynthesis (BBN) and the Reheating Phase After Inflation (RPAI), respectively. Next we have have found neutrino two-point function and shows a closed form decoupling of the hard ultraviolet (UV) divergent term from softened ultraviolet/infrared (UV/IR) mixing term and from the finite terms as well. For a certain choice of the noncommutative parameter $\theta$ which preserves unitarity, problematic UV divergent and UV/IR mixing terms vanish.
Non-perturbative modifications of the neutrino dispersion relations are assymptotically independent of the scale of noncommutativity in both the low and high energy  limits and may allow superluminal propagation.
\end{abstract}


\maketitle

\section{Introduction}

String theory indicated that noncommutative gauge field theory (NCGFT)
could be one of its low-energy effective theories \cite{Seiberg:1999vs}.
Studies on noncommutative particle phenomenology
\cite{Hinchliffe:2002km,Trampetic:2008bk} was motivated to find possible experimental signatures and/or predict/estimate bounds on space-time noncommutativity from collider physics experimental data: for example from the Standard Model (SM) invisible part of $Z \to \bar\nu\nu$ decays, and more important from the ultra high energy (UHE) processes occurring in the framework of the cosmic-ray neutrino physics. Constraint on the scale of the NCGFT, $\Lambda_{\rm NC}$, is possible due to a direct coupling of neutrinos to photons.

Significant progress has been obtained in the so-called Seiberg-Witten (SW) maps~\cite{Seiberg:1999vs} and enveloping algebra based models where one could deform commutative gauge theories with arbitrary gauge group and representation~
\cite{Madore:2000en,Jurco:2000ja,Jurco:2000fb,Jurco:2001my,Jurco:2001rq,Jackiw:2001jb,Calmet:2001na}. 
In our construction the noncommutative fields are 
obtained via SW maps from the original commutative fields.
It is commutative instead of the noncommutative gauge symmetry that is
preserved as the fundamental symmetry of the theory.
The constraints on the $\rm U_\star(1)$ charges, stated as ``no-go theorem'' \cite{Chaichian:2009uw}, are also rescinded  in our approach
\cite{Horvat:2011qn}, and the noncommutative extensions of particle physics covariant SM (NCSM) and the noncommutative grand unified theories (NCGUT) models \cite{Calmet:2001na,Horvat:2011qn,Behr:2002wx, Deshpande:2001mu,Duplancic:2003hg,Aschieri:2002mc,Melic:2005fm,Melic:2005am} were constructed. These allow a minimal deformation with no new particle content and with the sacrifice that interactions include infinitely many terms defined through recursion over the NC parameter $\theta^{\mu\nu}$; in practice cut-off at certain $\theta$-order.

In a simple model of NC spacetime local coordinates
$x^\mu$ are promoted to hermitian operators
$\hat x^\mu$ satisfying spacetime NC and implying 
uncertainty relations
\begin{equation}
[\hat x^\mu ,\hat x^\nu]=i\theta^{\mu\nu}\longrightarrow 
|\Delta x^\mu \Delta x^\nu| \geq \frac{1}{2}|\theta^{\mu\nu}|,
\label{x*x}
\end{equation}
where $\theta^{\mu\nu}$ is real, antisymmetric matrix. 
The Moyal-Weyl $\star$-product, relevant for the case of 
a constant $\theta^{\mu\nu}$, is defined as follows:
\begin{equation}
(f\star g)(x)=
e^{\frac{i}{2}h\frac{\partial}{{\partial x}^\mu}
\theta^{\mu\nu}
\frac{\partial}{{\partial y}^\nu}}f(x) g(y)\bigg|_{y\to x}.
\label{f*g}
\end{equation}
The operator commutation relation (\ref{x*x}) is then realized by  
the so-called $\star$-commutator
\begin{equation}
[\hat x^\mu ,\hat x^\nu]=[x^\mu \stackrel{\star}{,} x^\nu]=i\theta^{\mu\nu}.
\end{equation}

The perturbative quantization of noncommutative field theories was first
proposed in a pioneering paper by Filk \cite{Filk:1996dm}.
Other famous examples are the running of the coupling constant of NC QED \cite{MS-R} and the UV/IR mixing \cite{Minwalla:1999px,Matusis:2000jf}. Later well behaving one-loop quantum corrections to noncommutative scalar 
$\phi^4$ theories \cite{Grosse:2004yu,Magnen:2008pd,arXiv:1111.5553} 
and the NC QED \cite{Vilar:2009er} have been found. 
Also the SW expanded NCSM \cite{Calmet:2001na,Behr:2002wx,Duplancic:2003hg,Melic:2005fm}  at first order in $\theta$, albeit breaking Lorentz symmetry is anomaly free \cite{Martin:2002nr,Brandt:2003fx}, and has well-behaved one-loop quantum corrections \cite{Minwalla:1999px,Matusis:2000jf,MS-R,Bichl:2001cq,Buric:2006wm,Latas:2007eu,Martin:2007wv,Buric:2007ix,Martin:2009vg,Martin:2009sg,Tamarit:2009iy,Buric:2010wd}.
However, despite of some significant progress in the
models \cite{Grosse:2004yu,Magnen:2008pd,arXiv:1111.5553,Vilar:2009er,Bichl:2001cq,Martin:2002nr,Brandt:2003fx,Buric:2006wm,Latas:2007eu,Martin:2007wv,Buric:2007ix,Martin:2009vg,Martin:2009sg,Tamarit:2009iy,Buric:2010wd}, a better understanding of various models quantum loop corrections still remains in general a challenging open question. This fact is particularly true for the models constructed by using SW  map expansion in the NC parameter $\theta$, \cite{Madore:2000en,Calmet:2001na,Aschieri:2002mc,Schupp:2002up,Minkowski:2003jg}. Resulting models are very useful as effective field theories including their one-loop quantum properties \cite{Bichl:2001cq,Martin:2002nr,Brandt:2003fx,Buric:2006wm,Latas:2007eu,Martin:2007wv,Buric:2007ix,Martin:2009sg,Martin:2009vg,Tamarit:2009iy,Buric:2010wd} and relevant phenomenology \cite{Melic:2005hb,Ohl:2004tn,Alboteanu:2006hh,Ohl:2010zf,Melic:2005su,Tamarit:2008vy,Horvat:2009cm,Buric:2007qx}.

Discussions on the C,P,T, and CP properties of the noncommutative interactions are given in \cite{Melic:2005hb}, and in particular in \cite{Tamarit:2008vy}.  For example, 
fixing $\theta$ spontaneously breaks C, P, and/or CP discrete symmetries \cite{Aschieri:2002mc}. A breaking of C symmetry occurs in $Z\to \gamma\gamma$ process. One common approximation in those existing works is that only  the vertices linear in terms of the NC parameter $\theta$ were used. 
 
Recently, $\theta$-exact SW map and enveloping algebra based theoretical models were constructed in the framework of covariant noncommutative quantum gauge field theory \cite{Jackiw:2001jb}, and applied in loop computation 
\cite{Zeiner:2007,Schupp:2008fs,arXiv:1109.2485,arXiv:1111.4951}
and to the phenomenology, as well \cite{Horvat:2010sr,Horvat:2011iv}.

At $\theta$-order there are two important interactions that are suppressed  
and/or forbidden in the SM, the triple neutral gauge boson
\cite{Behr:2002wx,Duplancic:2003hg,Melic:2005fm},  
and the tree level coupling of neutrinos with photons \cite{Schupp:2002up, Minkowski:2003jg}, respectively. Here an expansion and cut-off in powers of 
the NC parameters $\theta^{\mu\nu}$ corresponds to an expansion in momenta  and restrict the range of validity to energies well below the NC scale $\Lambda_{\rm NC}$. Usually, this is no problem for experimental
predictions because the lower bound on the NC parameters
$\theta^{\mu\nu}=c^{\mu\nu}/\Lambda_{\rm NC}^2$ (the coefficients
$c^{\mu\nu}$ running between zero and one)
runs higher than typical momenta involved in a particular process.
However, there are exotic processes in the early universe as well as those involving ultra high energy cosmic rays
\cite{Horvat:2009cm,Horvat:2010sr,Horvat:2011iv,Horvat:2011wh}
in which the typical energy involved is higher than the current experimental
bound on the NC scale $\Lambda_{\rm NC}$.
Thus, the previous $\theta$-cut-off approximate results are inapplicable.
To cure the cut-off approximation, we are using $\theta$-exact
expressions, inspired by exact formulas for the SW
map \cite{Jurco:2001my,Okawa:2001mv,Martin:2012aw}, and expand in powers of gauge fields, as we did in \cite{Horvat:2011iv}. In $\theta$-exact models we have studied the UV/IR mixing 
\cite{Schupp:2008fs,arXiv:1109.2485}, the neutrino propagation 
\cite{arXiv:1111.4951} and also some NC photon-neutrino phenomenology 
\cite{Horvat:2009cm,Horvat:2010sr,Horvat:2011iv,Horvat:2011wh}, respectively.  
Due to the presence of the UV/IR mixing the $\theta$-exact model 
is not perturbatively renormalizable,  thus the relations of quantum corrections to the observations \cite{Horvat:2010km} are not entirely clear. 

In this work we present NCSM extended neutrino gauge bosons actions to all orders of  $\theta$ and study quantum properties: Neutrino two point function. Finally we discuss the decay width $\Gamma(Z\to \nu\nu)$ as functions of the NC scale $\Lambda_{\rm NC}$ for light-like noncommutativity which are allowed by unitarity condition \cite{Gomis:2000zz,Aharony:2000gz}.\\

\section{UHE cosmic ray motivation}

\begin{figure}[top]
\begin{center}
\includegraphics[width=8.5cm,height=5cm]{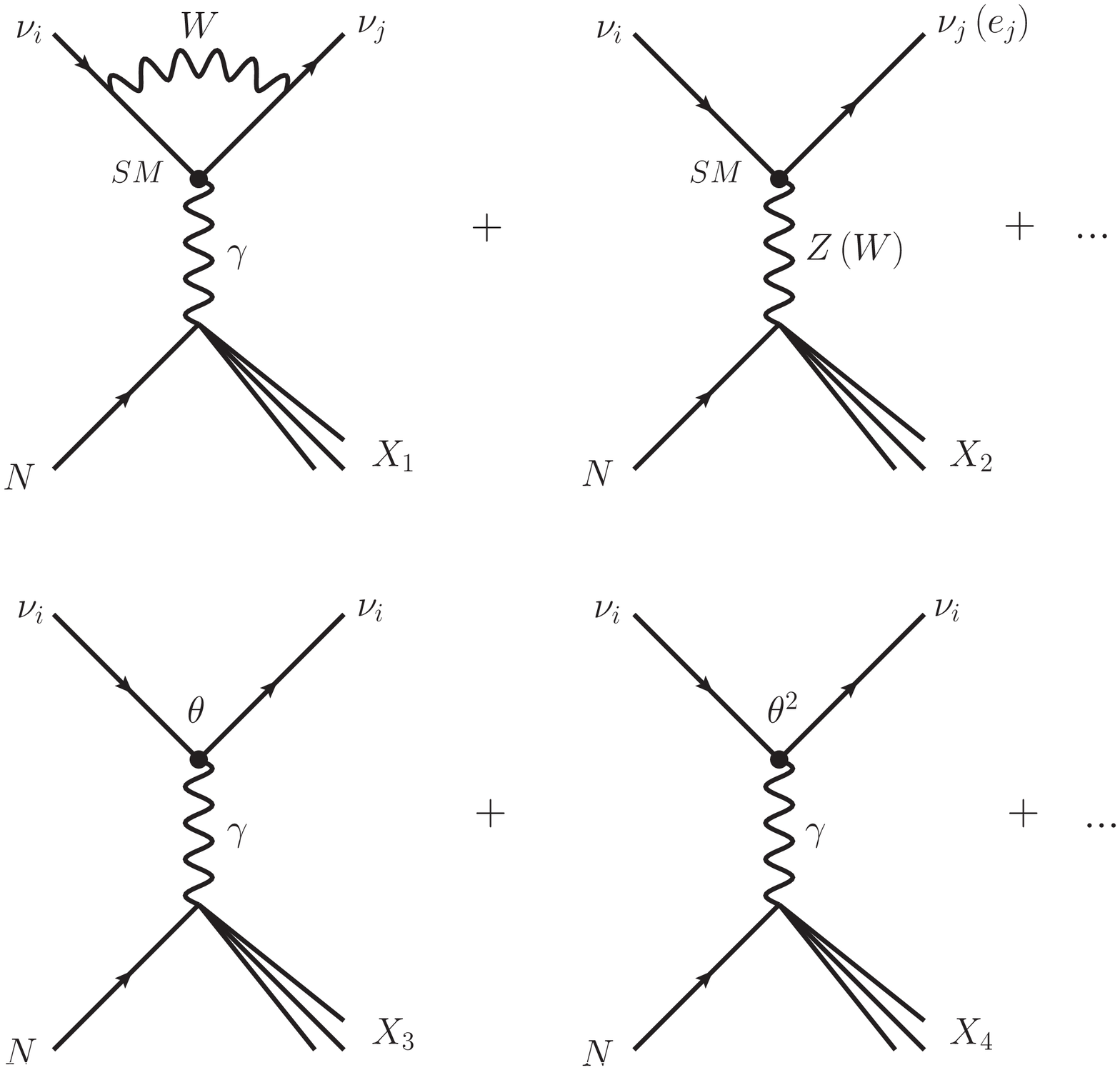}
\end{center}
\caption{Diagrams contributing to $\nu N\to\nu+X$ processes. }
\label{fig:trampslika}
\end{figure}

Direct coupling of gauge bosons to neutral  and ``chiral'' fermion particles \cite{Schupp:2002up,Horvat:2010sr,Horvat:2011iv}, via $\star$-commutator in the NC background, which plays the role of an external field in the theory, allow us to estimate a constraint on the scale of the noncommutative guge field theory, $\Lambda_{\rm NC}$, arising from ultra-high energy cosmic ray experiments involving $\nu$-nucleon inelastic cross section, see i.e. Fig. \ref{fig:trampslika}.
\begin{figure}[top]
\begin{center}
\includegraphics[width=8.5cm,height=5cm]{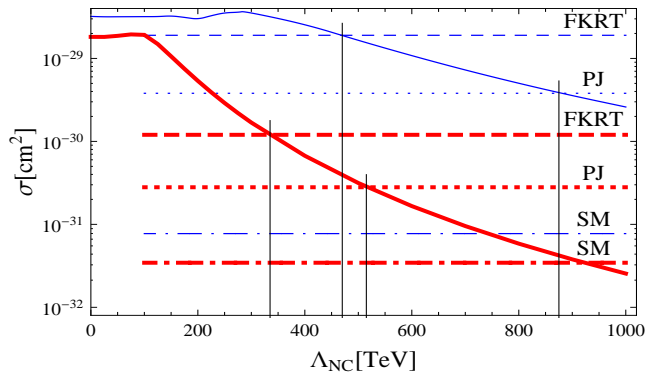}
\end{center}
\caption{$\nu N\to\nu\,+\, anything$ cross sections
vs. $\Lambda_{\rm NC}$ for $E_\nu=10^{10}~\GeV$ (thick lines) 
and $E_\nu=10^{11}~\GeV$ (thin lines). FKRT and PJ lines are the upper
bounds on the $\nu$-nucleon inelastic cross section, denoting different 
estimates for the cosmogenic $\nu$-flux. SM denotes the SM total
(charged current plus neutral current) $\nu$-nucleon inelastic cross
section. The vertical lines denote the intersections of our curves with the
RICE results.  }
\label{fig:ncSM-CrossSections}
\end{figure}
\begin{figure}[top]
\begin{center}
\includegraphics[width=8cm,height=5cm]{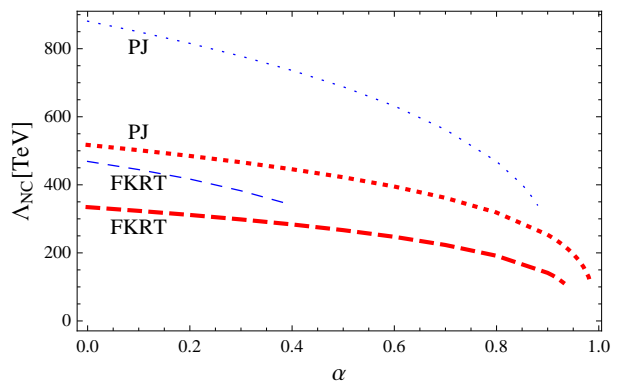}
\end{center}
\caption{The intersections of our curves with the RICE results (cf. Fig.2)
as a function of the fraction of Fe nuclei in the UHE cosmic rays. The
terminal point on each curve represents the highest fraction of Fe nuclei
above which no useful information on $\Lambda_{\rm NC}$ can be inferred with
our method.}
\label{fig:ncSM-LambdaVsAlpha}
\end{figure}

The observation of ultra-high energy (UHE) $\nu$'s from extraterrestrial
sources would open a new window to look to the cosmos, as such $\nu$'s may
easily escape very dense material backgrounds around local astrophysical
objects, giving thereby information on regions that are otherwise hidden to
any other means of exploration. In addition, $\nu$'s are not deflected on
their way to the earth by various magnetic fields, pointing thus back to
the direction of distant UHE cosmic-ray source candidates. This could also
help resolving the underlying acceleration in astrophysical sources.
     
In the energy spectrum of UHE cosmic rays at $\sim 4 \times 10^{19}$ eV the GZK-structure has been observed recently with high statistical accuracy \cite{Aglietta:2007yx}. Thus the flux of the so-called 
cosmogenic $\nu$'s, arising from photo-pion production on the
cosmic microwave background $p \gamma_{CMB} \rightarrow \Delta^{*}
\rightarrow N\pi$ and subsequent pion decay, is now guaranteed to exist. 
Possible ranges for the size of the flux of cosmogenic $\nu$'s can be obtained from separate analysis of the data from various large-scale observatories \cite{Fodor:2003ph,Protheroe:1995ft}. 

Note that there is the uncertainty in the flux of cosmogenic $\nu$'s regarding the
chemical composition of UHE cosmic rays (for details see \cite{Horvat:2010sr}). 
\begin{figure}
\begin{center}
\includegraphics[width=8cm,height=5.5cm]{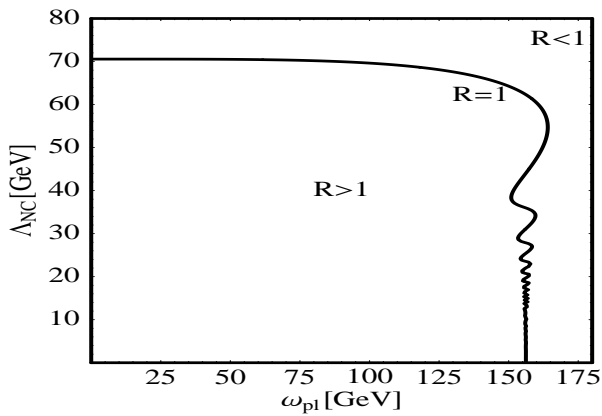}
\end{center}
\caption{The plot of scale $\Lambda_{\rm NC}$ versus the plasmon frequency 
$ \omega_{\rm pl}$ with $R=1$, from the plasmon decay into neutrino pairs 
$\gamma_{\rm pl} \to \bar\nu\nu$.}
\label{fig:RvsOmegaL}
\end{figure}
\begin{figure}
\begin{center}
\includegraphics[width=8cm,height=5.5cm]{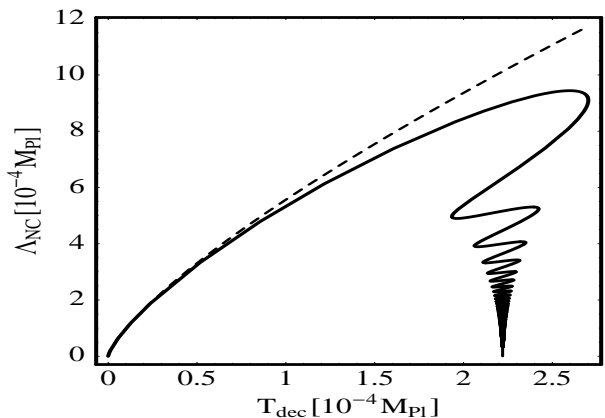}
\end{center}
\caption{The plot of the scale $\Lambda_{\rm NC}$ versus $T_{dec}$
for perturbative/exact solution ({dashed/full curve}).}
\label{fig:LambdaVsT}
\end{figure}
Usinging the upper bound on the $\nu N$ cross section derived from the RICE
Collaboration search results \cite{Kravchenko:2002mm} at $E_{\nu} =
10^{11}$ GeV ($4\times 10^{-3}$ mb for the FKRT $\nu$-flux \cite{Fodor:2003ph})), 
one can infer from $\theta$-truncated model    
on the NC scale $\Lambda_{\rm NC}$ to be greater than 455 TeV, 
a really strong bound. Here we have $\theta^{\mu\nu} \equiv c^{\mu\nu}/\Lambda_{\rm NC}^2 $ such that the matrix elements of $c$ are of order one.
One should however be
careful and suspect this result as it has been obtained from the conjecture
that the $\theta$-expansion stays well-defined in the kinematical region of interest. 
Although a heuristic criterion for the validity of the perturbative $\theta$-expansion,
$\sqrt{s}/\Lambda_{\rm NC} \lesssim\;1$, with $s = 2 E_{\nu}M_N$, would
underpin our result on $\Lambda_{\rm NC}$, a more thorough inspection on
the kinematics of the process does reveal a  more stronger energy
dependence  $E_{\nu}^{1/2} s^{1/4}/ \Lambda_{\rm NC} \lesssim 1$.
In spite of an additional phase-space suppression for small $x$'s in  
the $\theta^2$-contribution \cite{Ohl:2004tn} of
the cross section relative to the $\theta$-contribution, we find an unacceptably 
large ratio $\sigma({\theta^2})/\sigma({\theta}) \simeq 10^4$,
at $\Lambda_{\rm NC}=455$ TeV.
Hence, the bound on $\Lambda_{\rm NC}$ obtained this way is incorrect, 
and our last resort 
is to modify the model adequately to include the full-$\theta$
resummation, thereby allowing us to compute nonperturbatively in $\theta$.

Total cross section, as a function of the NC scale at fixed $E_{\nu} = 10^{10}$ GeV  and $E_{\nu} = 10^{11}$ GeV, together with the upper bounds depending on the actual size of the cosmogenic $\nu$-flux (FKRT \cite{Fodor:2003ph} and PJ \cite{Protheroe:1995ft}) as well as the total SM cross sections at these energies, are
depicted in our Figure \ref{fig:ncSM-CrossSections}. In order to maximize 
the NC $\theta$-exact effect we choose $c_{01}-c_{13}=c_{02}-c_{23}=c_{03}=1$.

Even if the future data confirm that UHE cosmic rays are composed mainly of Fe
nuclei, as indicated by the PAO data, then still
valuable information on $\Lambda_{\rm NC}$ can be obtained with our method, as
seen in Fig.\ref{fig:ncSM-LambdaVsAlpha}. 
Here we see the intersections of our curves with the
RICE results (cf. Fig.\ref{fig:ncSM-CrossSections})
as a function of the fraction $\alpha$ of Fe nuclei in the UHE cosmic rays.
On top of results, presented in 
Figs.\ref{fig:ncSM-CrossSections} and \ref{fig:ncSM-LambdaVsAlpha},
we also have the NC scale given as a function of 
the plasmon frequency, from the plasmon decay into neutrino pairs
$\gamma_{\rm pl} \to \bar\nu\nu$ (Fig.\ref{fig:RvsOmegaL}), and as a function of 
the $T_{dec}$ from BBN (Fig.\ref{fig:LambdaVsT}), respectively. 
All results depicted in Figs.\ref{fig:ncSM-CrossSections}-\ref{fig:LambdaVsT},
shows convergent behavior.  In our opinion those were the strong signs   
to continue research towards  quantum properties and phenomenology of such $\theta$-exact noncommutative gauge field theory model. 

\section{Consistency of the SW map and enveloping algebra approach to NCGFT}

The choice of gauge group appears to be severely restricted in a noncommutative setting \cite{Seiberg:1999vs}: The star commutator of two Lie algebra valued gauge fields will involve the anti-commutator as well as the commutator of the Lie algebra generators. The algebra still closes for Hermitian matrices, but it is for instance not possible to impose the trace to be zero. This observation can be interpreted in two ways:\\
(a) The choice of gauge group is restricted to $\rm U(N)$ in the fundamental, anti-fundamental or adjoint representation; or\\
(b) the gauge fields are valued in the enveloping algebra of a Lie algebra and then any (unitary) representation is possible.\\
The case (a) applies also to the $\rm U(1)$ case and imposes severe restrictions on the allowed charges; it has been studied carefully and has led to ``theorems''~\cite{Hayakawa:1999yt,Chaichian:2001mu}. The second case avoids the restrictions on the gauge group and choice of representation, but needs to address the potential problem of too many degrees of freedom, since all coefficient functions of the monomials in the generators could a priori be physical fields. The solution to this problem is that the coefficient fields are not all independent. They are rather functions of the correct number of ordinary gauge fields via Seiberg-Witten maps and their generalizations. The situation is reminiscent of the construction of superfields and supersymmetric actions in terms of ordinary fields in supersymmetry. This method, referred as Seiberg-Witten map or enveloping algebra approach avoids both the gauge group and the U(1) charge issues. It was shown mathematically rigorously that any U(1) gauge theory on an arbitrary Poisson manifold can be deformation-quantized to a noncommutative gauge theory via the the enveloping algebra approach~\cite{Jurco:2001kp} and later extended to the non-Abelian gauge groups~\cite{2007arXiv0711.2965B,2009arXiv0909.4259B}. The important step that has been missed in a paper 
\cite{Chaichian:2009uw} opposing above conclusions, is the use of reducible representations \cite{Horvat:2011qn}.

Following \cite{Horvat:2011qn} we introduce a consistent
noncommutative, Seiberg-Witten map and enveloping algebra based theory:
Let $\hat \Phi [\Phi, A_\mu]$, $\hat A_\mu[A_\mu]$, $\hat \Lambda [\Lambda, A_\mu]$ be the SW map expanded fields (consider for example the well-known non-abelian maps for the Moyal-Weyl case~\cite{Seiberg:1999vs}).
Under an ordinary gauge transformation $\delta$ of the underlying fields
$\phi_i(x)$, $i=1,2,3$ and $a_\mu$ the SW expanded fields transform like
it is expected for noncommutative fields.

Since in the noncommutative case the order of fields matters, there are in fact more choices than the one given in (4). In general all fields carry left and right charges that combines into the total commutative charge. Gauge invariance requires that the respective charges of neighboring fields must match with opposite signs. In the notation of (2) and (4), we have:
\begin{equation}
\delta \hat \Phi = i \hat \Lambda^{L}  \star \hat \Phi - i \hat \Phi \star \hat \Lambda^{R}  \,.
\end{equation}
Using the associativity of the star product one can easily verify the formal consistency relation
\begin{equation}
[\delta_{\hat\Lambda},\delta_{\hat\Sigma}]\hat\Phi=[i\hat \Lambda^{L} \stackrel{\star}{,}i\hat \Sigma^{L} ]\star\hat\Phi-\hat\Phi\star[i\hat \Lambda^{R} \stackrel{\star}{,}i\hat\Sigma^{R}].
\label{hybridconsistency}
\end{equation}
Therefore the noncommutative gauge transformations $\hat \Lambda^{L/R}$ can be constructed from the classical fields and parameters $A_\mu^{L/R} = a_\mu(x) Q^{L/R}$ and $\Lambda^{L/R} = \lambda(x)  Q^{L/R}$ with
$Q^{L/R} = \text{diag}(q_1^{L/R},q_2^{L/R},q_3^{L/R})$ and $q_i = q_i^{L} - q_i^{R}$ by so-called hybrid Seiberg-Witten  maps \cite{Calmet:2001na,Schupp:2001we}.
The hybrid covariant derivative is given by $\hat D_\mu \hat\Phi=\partial_\mu\hat\Phi-i \hat A_\mu^{L}\star\hat\Phi
+i\hat\Phi\star \hat A_\mu^{R}$. Thanks to \eqref{hybridconsistency} the left and right NC gauge fields $\hat A_\mu^ {L/R}$ are constructed from $A_\mu^{L/R}$ only, respectively. The gauge field action could be written as
\begin{equation}
\mathcal L_{gauge}=-\frac{1}{4g^2}\tr\left(\hat F_{\mu\nu}^{L} \star \hat F^{\mu\nu L}+\hat F_{\mu\nu}^{R} \star \hat F^{\mu\nu R}\right)\,,
\label{FLR}
\end{equation}
with $g:=e\sqrt{\tr {(Q^{L})}^2+\tr {(Q^{R})}^2}$.
In \cite{Horvat:2011qn} we have employed this constructon on deformed Yukawa couplings. Namely,  in the Yukawa terms, a star product deformation would prevent the charge summation. The hybrid SW map \cite{Calmet:2001na, Schupp:2001we} is introduced to recover gauge invariance. Thus the classical charge $q$ is split into left and right charges $q=q^{L}-q^{R}$, as we have seen above.

\section{Covariant $\theta$-exact ${\rm U_{\star}(1)}$ model }

We start with the following SW type of NC $\rm U_{\star}(1)$ gauge model:
\begin{equation}
S=\int-\frac{1}{4}F^{\mu\nu}\star F_{\mu\nu}+i\bar\Psi\star\fmslash{D}\Psi\,,
\label{S}
\end{equation}
with the NC definitions of the nonabelian field strength and the covariant derivative, respectively:
\begin{eqnarray}
F_{\mu\nu}&=&\partial_\mu A_\nu-\partial_\nu
A_\mu-i[A_\mu\stackrel{\star}{,}A_\nu],
\nonumber\\
D_\mu\Psi&=&\partial_\mu\Psi-i[A_\mu\stackrel{\star}{,}\Psi].
\label{DF}
\end{eqnarray}
All noncommutative fields in this action $(A_\mu,\Psi)$ are images under (hybrid) Seiberg-Witten maps
of the corresponding commutative fields $(a_\mu,\psi)$.
Here we shall interpret the NC fields as valued in
the enveloping algebra of the underlying gauge group.
This naturally corresponds to an expansion in powers of
the gauge field $a_\mu$ and hence in powers of
the coupling constant $e$. At each order in $a_\mu$ we shall
determine $\theta$-exact expressions.

In the next step we expand the action in terms of
the commutative gauge parameter $\lambda$
and fields $a_\mu$ and $\psi$ using the SW map solution \cite{Schupp:2008fs}
up to the $\mathcal O(a^3)$ order:
\begin{eqnarray}
\Lambda&=&\lambda-\frac{1}{2}\theta^{ij}a_i\star_2\partial_j\lambda \,,
\nonumber\\
A_\mu&=&\,a_\mu-\frac{1}{2}\theta^{\nu\rho}{a_\nu}\star_2(\partial_\rho
a_\mu+f_{\rho\mu}),
\nonumber\\
\Psi&=&\psi-\theta^{\mu\nu}
{a_\mu}\star_2{\partial_\nu}\psi
\nonumber\\
&+&\frac{1}{2}\theta^{\mu\nu}\theta^{\rho\sigma}
\bigg[(a_\rho\star_2(\partial_\sigma
a_\mu+f_{\sigma\mu}))\star_2{\partial_\nu}\psi
\nonumber\\
&+&2a_\mu{\star_2}
(\partial_\nu(a_\rho{\star_2}\partial_\sigma\psi))
-a_\mu{\star_2}(\partial_\rho
a_\nu{\star_2}\partial_\sigma\psi)
\nonumber\\
&-&\big(a_\rho\partial_\mu\psi(\partial_\nu
a_\sigma+f_{\nu\sigma})
-\partial_\rho\partial_\mu\psi a_\nu
a_\sigma\big)_{\star_3}\bigg],
\label{SWmap}
\end{eqnarray}
with $\Lambda$ being the NC gauge parameter and
$f_{\mu\nu}$ is the abelian commutative field strength
$f_{\mu\nu}=\partial_\mu a_\nu-\partial_\nu a_\mu$.

The generalized Mojal-Weyl star products
$\star_2$ and $\star_3$, appearing in (\ref{SWmap}), are defined, respectively, as
\begin{eqnarray}
f(x)\star_2 g(x)&=&[f(x) \stackrel{\star}{,}g(x)]
\nonumber\\
&=&\frac{\sin\frac{\partial_1\theta
\partial_2}{2}}{\frac{\partial_1\theta
\partial_2}{2}}f(x_1)g(x_2)\bigg|_{x_1=x_2=x}\,,
\label{f*2g}
\end{eqnarray}
\begin{eqnarray}
(f(x)g(x)h(x))_{\star_3}&=&\Bigg(\frac{\sin(\frac{\partial_2\theta
\partial_3}{2})\sin(\frac{\partial_1\theta(\partial_2+\partial_3)}{2})}
{\frac{(\partial_1+\partial_2)\theta \partial_3}{2}
\frac{\partial_1\theta(\partial_2+\partial_3)}{2}}
\nonumber\\
&+&\{1\leftrightarrow 2\}\,\Bigg)f(x_1)g(x_2)h(x_3)\bigg|_{x_i=x},
\nonumber\\
\label{fgh*3}
\end{eqnarray}
where $\star$ is associative but noncommutative, while $\star_2$ and $\star_3$
are both commutative but nonassociative. 

\begin{figure}
\begin{center}
\includegraphics[width=8cm,height=4cm]{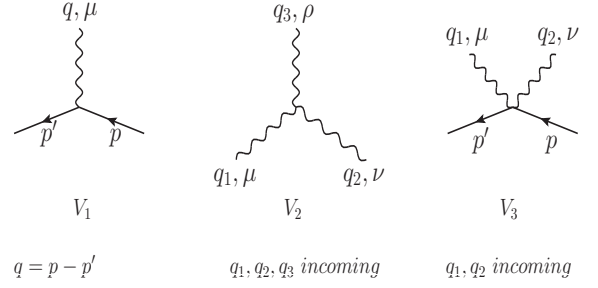}
\end{center}
\caption{Three- and your-field vertices}
\label{fig:Bayzell11Vert}
\end{figure}

The resulting expansion defines $\theta$-exact neutrino-photon ${\rm U_{\star}(1)}$ actions, for a gauge and a matter sectors respectively. Pure gauge field (3-photon) action reads:
\begin{eqnarray}
S_g&=&\int \;i\partial_\mu a_\nu\star[a^\mu\stackrel{\star}{,}a^\nu]
\nonumber\\
&+&\frac{1}{2}\partial_\mu
\bigg(\theta^{\rho\sigma}a_\rho\star_2(\partial_\sigma a_{\nu}+f_{\sigma\nu})\bigg)\star
f^{\mu\nu}.
\label{Sgauge}
\end{eqnarray}
The photon-fermion action up to 2-photon 2-neutrino fields can be derived
by using the first order gauge field and the second order neutrino field expansions,
\begin{eqnarray}
S_f&=&\int \;\bigg(\bar\psi
+(\theta^{ij}\partial_i\bar\psi \star_2
a_j)\bigg)\gamma^\mu[a_\mu\stackrel{\star}{,}\psi]
\nonumber\\
&+&
i(\theta^{ij}\partial_i\bar\psi
\star_2 a_j)\fmslash\partial\psi-i\bar\psi\star
\fmslash\partial(\theta^{ij}
a_i\star_2\partial_j\psi)
\nonumber\\
&-&
\!\bar\psi\gamma^\mu[a_\mu\!\stackrel{\star}{,}\!\theta^{ij}
a_i\!\star_2\!\partial_j\psi]\!
\nonumber\\
&-&
\!\bar\psi\gamma^\mu
\bigg[\frac{1}{2}\theta^{ij}a_i\!\star_2\!(\partial_j
a_\mu\!+\!f_{j\mu})\!\stackrel{\star}{,}\!\psi\bigg]\!
\nonumber\\
&-&
\!i(\theta^{ij}\partial_i\bar\psi
\!\star_2\!a_j)\fmslash\partial(\theta^{kl}
a_k\!\star_2\!\partial_l\psi)
\nonumber\\
&+&
\frac{i}{2}\theta^{ij}\theta^{kl}
\bigg[(a_k\star_2(\partial_l
a_i+f_{li}))\star_2\partial_j\bar\psi
\nonumber\\
&+&
2a_i\star_2(\partial_j(a_k\star_2\partial_l\bar\psi))-
a_i\star_2(\partial_k
a_j\star_2\partial_l\bar\psi)
\nonumber\\
&+&
\big(a_i\partial_k\bar\psi(\partial_j
a_l+f_{jl})-\partial_k\partial_i\bar\psi a_j a_l\big)_{\star_3}\bigg]
\fmslash\partial\psi
\nonumber\\
&+&
\frac{i}{2}\theta^{ij}\theta^{kl}\bar\psi\fmslash\partial
\bigg[(a_k\star_2(\partial_l
a_i+f_{li}))\star_2\partial_j\psi
\nonumber\\
&+&
\!2a_i\!\star_2\!
(\partial_j(a_k\!\star_2\!\partial_l\psi))\!-\!a_i\!\star_2\!(\partial_k
a_j\!\star_2\!\partial_l\psi)\!
\nonumber\\
&+&
\big(a_i\partial_k\psi(\partial_j
a_l\!+\!f_{jl})\!-\!\partial_k\partial_i\psi a_j
a_l\big)_{\star_3}\bigg]\,.
\label{Sfermion}
\end{eqnarray}
Note that actions for gauge and matter fields obtained above,
(\ref{Sgauge}) and (\ref{Sfermion}) respectively, are nonlocal
objects due to the presence of the star products: $\star$, $\star_2$ and $\star_3$.
Feynman rules from above actions, represented in Fig.\ref{fig:Bayzell11Vert},
are given explicitly in \cite{arXiv:1111.4951}.
\\

\section{Quantum properties: Neutrino two-point function}

\begin{figure}
\begin{center}
\includegraphics[width=8cm,height=6cm]{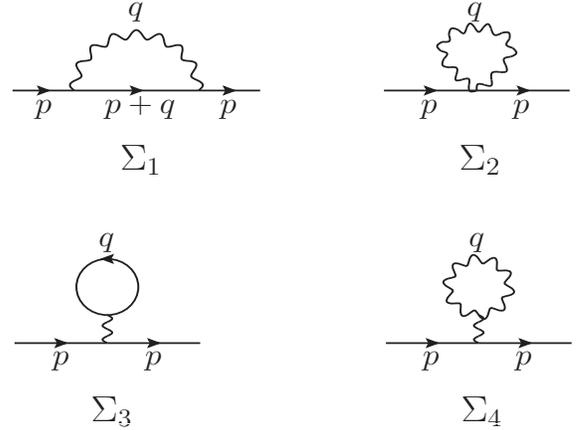}
\end{center}
\caption{One-loop self-energy of a massless neutrino}
\label{Sigma1-loop}
\end{figure}

As depicted in Fig. \ref{Sigma1-loop}, there are four Feynman diagrams
contributing  to the  $\nu$-self-energy at one-loop.
With the aid of (\ref{Sfermion}), 
we have verified by explicit calculation that the 4-field
tadpole ($\Sigma_2$) does vanish. The 3-fields tadpoles ($\Sigma_3$ 
and $\Sigma_4$) can be ruled out by invoking
the NC charge conjugation symmetry \cite{Aschieri:2002mc}.
Thus only the $\Sigma_1$ diagram needs to be evaluated.   
In spacetime of the dimensionality $D$ we obtain
\begin{eqnarray}
\Sigma_1
&=&\mu^{4-D}\int \frac{d^D q}{(2\pi)^D}
\bigg(\frac{\sin\frac{q\theta p}{2}}{\frac{q\theta
p}{2}}\bigg)^2\frac{1}{q^2}\frac{1}{(p+q)^2}
\nonumber\\
&\cdot & \Bigg[(q\theta p)^2(4-D)(\fmslash p+\fmslash q)
\label{Sigma1}\\
&+&(q\theta p)\bigg(/\!\!\tilde q(2p^2+2p\cdot q)-/\!\!\tilde
p(2q^2+2p\cdot q)\bigg)
\nonumber\\
&+&\bigg(\fmslash p(\tilde q^2(p^2+2p\cdot q)-q^2(\tilde p^2+2\tilde
p\cdot\tilde q) )
\nonumber\\
&+&
\fmslash q(\tilde p^2(q^2+2p\cdot q)-p^2(\tilde
q^2+2\tilde p\cdot\tilde q ))\bigg)\Bigg]\,,
\nonumber
\end{eqnarray}
where ${\tilde p}^\mu=(\theta p)^\mu=\theta^{\mu\nu} p_\nu $,
and in addition ${\tilde{\tilde p}}^\mu=(\theta\theta p)^\mu
=\theta^{\mu\nu}\theta_{\nu\rho}p^\rho$.
To perform computations of those integrals using the dimensional
regularization method, we first use
the Feynman parametrization on the quadratic denominators,
then the Heavy Quark Effective theory (HQET)
parametrization \cite{Grozin:2000cm} is used to combine
the quadratic and linear denominators.
In the next stage we use the Schwinger
parametrization to turn the denominators
into Gaussian integrals. Evaluating the relevant integrals for
$D=4-\epsilon$ in the limit $\epsilon\to 0$, we obtain the 
closed form expression for the self-energy
\begin{eqnarray}
\Sigma_1&=&\gamma_{\mu}
\bigg[p^{\mu}\: A+(\theta{\theta p})^{\mu}\;\frac{p^2}{(\theta p)^2}\;B\bigg]\,,
\label{sigma1AB}\\
A&=& \frac{-1}{(4\pi)^2}
\bigg[p^2\;\bigg(\frac{\tr\theta\theta}{(\theta p)^2}
+2\frac{(\theta\theta p)^2}{(\theta p)^4}\bigg) A_1 
\nonumber\\
&+&
 \bigg(1+p^2\;\bigg(\frac{\tr\theta\theta}{(\theta p)^2}
+\frac{(\theta\theta p)^2}{(\theta p)^4}\bigg)\bigg)A_2 \bigg]\,,
\label{A}\\
A_1&=&\frac{2}{\epsilon}+\ln(\mu^2(\theta p)^2)
+ \ln({\pi e^{\gamma_{\rm E}}})
\label{A1}\\
&+&
\sum\limits_{k=1}^\infty \frac{\left(p^2(\theta p)^2/4\right)^k}{\Gamma(2k+2)}
\left(\ln\frac{p^2(\theta p)^2}{4} + 2\psi_0(2k+2)\right) \,,
\nonumber\\
A_2&=&-\frac{(4\pi)^2}{2} B = -2
\nonumber\\
&+& \sum\limits_{k=0}^\infty
\frac{\left(p^2(\theta p)^2/4\right)^{k+1}}{(2k+1)(2k+3)\Gamma(2k+2)}
\Bigg(\ln\frac{p^2(\theta p)^2}{4} 
\nonumber\\
&-&
2\psi_0(2k+2)
- \frac{8(k+1)}{(2k+1)(2k+3)} \Bigg),
\label{A2}
\end{eqnarray}
with  $\gamma_{\rm E}\simeq0.577216$ being Euler's constant.

The $1/\epsilon$ UV divergence
could in principle be removed by a properly chosen counterterm.
However due to the specific momentum-dependent 
coefficient in front of it, a nonlocal form for it is required. 

\subsection{{\rm \bf UV/IR} mixing}

Turning to the UV/IR mixing problem, we recognize
a soft UV/IR mixing term represented by a logarithm,
\begin{equation}
\Sigma_{\rm UV/IR}={\fmslash p} \:\frac{p^2}{(4\pi)^2}
\left(\ln\frac{1}{\mu^2(\theta p)^2}\right)
\bigg(\frac{\tr\theta\theta}{(\theta p)^2}
+2\frac{(\theta\theta p)^2}{(\theta p)^4}\bigg).
\label{lnUV/IR}
\end{equation}
Instead of dealing with nonlocal counterterms, we take a different route
here to  cope with various divergences besetting (\ref{sigma1AB}). Since $\theta^{0i}
\neq 0$ makes a NC theory nonunitary \cite{Gomis:2000zz},
we can, without loss of generality,
chose $\theta$ to lie in the (1, 2) plane
\begin{equation}
\theta^{\mu\nu}=\frac{1}{\Lambda_{\rm NC}^2}
\begin{pmatrix}
0&0&0&0\\
0&0&1&0\\
0&-1&0&0\\
0&0&0&0
\end{pmatrix}.
\label{degen}
\end{equation}
Automatically, this produces
\begin{equation}
\frac{\tr\theta\theta}{(\theta p)^2}
+2\frac{(\theta\theta p)^2}{(\theta p)^4}=0,\:\forall p.
\label{degen1}
\end{equation}
With (\ref{degen1}), $\Sigma_1$,
in terms of Euclidean momenta, receives the following form:
\begin{equation}
\Sigma_1=\frac{-1}{(4\pi)^2}\gamma_\mu\left[p^\mu
\bigg(1 + \frac{\tr\theta\theta}{2}\frac{p^2}{(\theta p)^2}\bigg)
-2(\theta\theta p)^\mu\frac{p^2}{(\theta p)^2} \right] A_2.
\label{Sigmabuble}
\end{equation}
By inspecting (\ref{A2}) one can be easily convinced that $A_2$ is 
free from the $1/\epsilon$ divergence and the UV/IR mixing term, being also 
well-behaved in the infrared, in the $\theta \rightarrow 0$ 
as well as $\theta p \rightarrow 0$
limit. We see, however, that the two terms in (\ref{Sigmabuble}), 
one being proportional to
$\fmslash{p}$ and the other proportional to $\fmslash{\tilde{\tilde p}}$,
are still ill-behaved in the $\theta p \rightarrow 0$
limit. If, for the choice (\ref{degen}), $P$ denotes the momentum in the (1, 2)
plane, then $\theta p = \theta P$. For instance, a particle moving
inside the NC plane with
momentum $P$ along the one axis, has a spatial extension of size $|\theta P|$
along the other. For the choice (\ref{degen}), $\theta p \rightarrow 0$ corresponds to a zero momentum projection onto the (1, 2) plane. Thus, albeit in our approach the commutative limit ($\theta \rightarrow 0$) is smooth at the quantum level,
the limit when an extended object (arising due to the fuzziness of space)
shrinks to zero, is not. We could surely claim that in our approach the
UV/IR mixing problem is considerably softened; on the other hand, we have
witnessed how the problem strikes back in an unexpected way. This is, at the
same time, the first example where this two limits are not degenerate.

\subsection{Neutrino dispersion relations}

In order to probe physical consequence of
the 1-loop quantum correction, with $\Sigma_{1-loop_{M}}$ 
from Eq. (3.25) in \cite{arXiv:1111.4951}, 
we consider the modified propagator
\begin{equation}
\frac{1}{\fmslash\Sigma}=\frac{1}{\fmslash p-\Sigma_{1-loop_{M}}}=
\frac{\fmslash\Sigma}{\Sigma^2}\,.
\label{disp}
\end{equation}
We further choose the NC parameter to be (\ref{degen})
so that the denominator is finite and can be expressed explicitly:
\begin{equation}
\hspace{-.05cm}\Sigma^2=p^2\left[\hat A_2^2\left(\frac{p^4}{p^4_r}
+2\frac{p^2}{p_r^2}+5\right)-\hat A_2\left(6+2\frac{p^2}{p_r^2}\right)+1\right],\,(23)
\nonumber
\end{equation}
where $p_r$ represents $r$-component of the momentum $p$
in a cylindrical spatial coordinate system and $\hat A_2= e^2A_2/(4\pi)^2=-B/2$.

From above one see that $p^2=0$ defines one set of the dispersion relation,
corresponding to the dispersion for the massless neutrino mode,
however the denominator $\Sigma^2$ has one more coefficient $\Sigma'$
which could also induce certain zero-points. Since the $\hat A_2$ is a
function of a single variable $p^2p_r^2$, with $
p^2=p^2_0-p^2_1-p^2_2-p^2_3\,\,\rm and\,\,p^2_r=p^2_1+p^2_2$,
the condition $\Sigma'=0$  can be expressed as a simple algebraic equation
\begin{equation}
\hat A_2^2z^2-2\left(A_2-\hat A_2^2\right)z
+\left(1-6\hat A_2+5\hat A_2^2\right)=0\,,
\label{A2z2}
\end{equation}
of new variables $z:=p^2/p_r^2$, in which the coefficients
are all functions of $y:=p^2p^2_r/\Lambda^4_{\rm NC}$.

The two formal solutions of the equation (\ref{A2z2})
\begin{equation}
z=\frac{1}{\hat A_2}\left[\left(1-\hat A_2\right)\pm\,2\left(\hat A_2
-\hat A_2^2\right)^{\frac{1}{2}}\right]\,,
\label{zsolutions}
\end{equation}
are birefringent.
The behavior of solutions (\ref{zsolutions}), is next analyzed at
two limits $y\to \,0$, and $y\to\,\infty$.

\subsubsection{The low-energy regime: $p^2p^2_r \ll\Lambda^4_{\rm NC}$}

For $y \ll 1$ we set $\hat A_2$ to its zeroth order value $ e^2/8\pi^2$, 
\begin{eqnarray}
p^2&\sim& \left(\left(\frac{8\pi^2}{e^2}-1\right)\pm 2
\left(\frac{8\pi^2}{e^2}-1\right)^{\frac{1}{2}}\right)\cdot\,p_r^2
\nonumber\\
&\simeq&\left(859\pm 59\right)\cdot\,p_r^2\,,
\label{859pm59}
\end{eqnarray}
obtaining two (approximate) zero points. From the definition of $p^2$
and $p_r^2$ we see that both solutions are real and positive.
Taking into account the higher order (in y) correction
these poles will locate nearby the real axis of
the complex $p_0$ plane thus correspond to some metastable modes with
the above defined dispersion relations. As we can see, 
the modified dispersion relation \eqref{859pm59} does not depend on 
the noncommutative scale, therefore it introduces a discontinuity in 
the $\Lambda_{\rm NC}\to\infty$ limit, 
which is not unfamiliar in noncommutative theories.

\subsubsection{The high-energy regime: $p^2p^2_r \gg \Lambda^4_{\rm NC}$}
At $y\gg 1$ we analyze the asymptotic behavior of
\begin{equation}
A_2\sim \frac{i\pi^2}{8} {\sqrt y}\left(1-\frac{16i}{\pi y}
e^{-\frac{i}{2}{\sqrt y}}\right)+\mathcal O\left(y^{-1}\right),
\end{equation}
from \cite{arXiv:1111.4951}, therefore \eqref{zsolutions} can be reduced to
\begin{equation}
z\sim -1\pm 2i \;\; \rightarrow \;\;p^2_0\sim p^2_3\pm 2i p^2_r.
\end{equation}
We thus reach two unstable deformed modes besides the usual mode $p^2=0$ in the high energy regime. Here again the leading order deformed dispersion relation does not depend on the noncommutative scale $\Lambda_{\rm NC}$.

\subsection{The alternative action self-energy}

Using the Feynman rule of the alternative action (2.15) from Ref \cite{arXiv:1111.4951}, which is a consequence of the SW freedom, we find the following contribution to the neutrino self-energy from diagram $\Sigma_1$
\begin{equation}
\Sigma_{1_{alt}}=\fmslash p\: \frac{8}{3}\frac{1}{(4\pi)^2} \frac{1}{{(\theta p)^2}}
\bigg(\frac{\tr\theta\theta}{(\theta p)^2}
+4\frac{(\theta\theta p)^2}{(\theta p)^4}\bigg)\,.
\label{A9}
\end{equation}
The detailed computation is presented in Appendix B of Ref. \cite{arXiv:1111.4951}. 
We notice that there are no hard
$1/\epsilon$ UV divergent and no logarithmic UV/IR mixing terms,
and the finite terms like in $A_1$ and $A_2$ are also absent.
Thus the subgraph $\Sigma_1$ for the alternative action (2.15) in \cite{arXiv:1111.4951} does not require any counter-term. However, the result (\ref{A9}), does express powerful UV/IR mixing effect, that is in terms of scales terms, the $\Sigma_{1_{alt}}$ experience
the forth-power of the {\it NC-scale/momentum-scale} ratios
$\sim |p|^{-2}|\theta p|^{-2}$ in (\ref{A9}), i.e. we are dealing with
the $\Sigma_{1_{alt}}\sim{\fmslash p}\left({\Lambda_{\rm NC}}/{p}\right)^4$
within the ultraviolet and infrared limits for $\Lambda_{\rm NC}$ and $p$, 
respectively.

\section{Phenomenology: $Z\to \nu\bar{\nu}$ decay rate }

To illustrate another phenomenologicall effects of our $\theta$-exact construction, 
we present a computation the $Z\to\nu\bar\nu$ decay rate in 
the Z--boson rest frame, which is then readily to be compared with 
the precision Z resonance measurements, where Z is almost at rest.
Since the complete $Z\nu\nu$ interaction on noncommutative spaces was discussed in details in \cite{Horvat:2011qn},  we shall not repeat it here. We only give the {\it almost complete} $Z\nu\bar\nu$ vertex from \cite{Horvat:2011qn}
\begin{equation}
\begin{split}
&\Gamma^\mu(p^\prime,p)=
i\frac{g}{2\cos\theta_W}
\Bigg(
\gamma^\mu
+\frac{i}{2} F_\bullet(p^\prime,p)
\\&
\cdot \bigg[ (p^\prime \theta p)\gamma^\mu + (\theta p')^{\mu} {\fmslash p}
  -(\theta p)^\mu {\fmslash p}^{\,\prime} \bigg]
  \Bigg)\frac{1-\gamma_5}{2}
\\&
+\frac{\kappa e}{2}\tan\theta_W F_{\star_2}(p^\prime,p)
\\&
\cdot \bigg[ (p^\prime \theta p)\gamma^\mu + (\theta p')^{\mu} {\fmslash p}
  -(\theta p)^\mu {\fmslash p}^{\,\prime} \bigg]~,
  \end{split}
\label{Z2fbarfNC:Vertex}
\end{equation}
where $\kappa$ is an arbitrary constant\footnote{The constant $\kappa$ measures a correction from the  $\star$-commutator coupling of the right handed neutrino $\nu_R$ to the noncommutative hypercharge $\rm U_{\star}(1)_Y$ gauge field $B^0_\mu[\kappa]$. Coupling is chiral blind and it vanishes in the commutative limit. The non-$\kappa$-proportional term, on the other hand, is the noncommutative deformation of standard model Z-neutrino coupling, which involves the left handed neutrinos only. Details can be found in section four of \cite{Horvat:2011qn}. }, and
\begin{eqnarray}
(p^\prime \theta p)F_\bullet(p^\prime,p)
&=& -2 i \left(
1-\exp\left(i\frac{M_Z\, p}{2 \Lambda _{\text{NC}}^2}\cos\vartheta\right)
\right)~,
\nonumber\\
(p^\prime \theta p)F_{\star_2}(p^\prime,p)
&=&
-2 \sin\left( \frac{M_Z\, p}{2 \Lambda _{\text{NC}}^2}\cos\vartheta \right)~.
\label{p'tpds}
\end{eqnarray}
Note here that due to the equations of motions, for massless on-shell neutrinos
the terms $[ (\theta p')^{\mu} {\fmslash p} -(\theta p)^\mu {\fmslash p'} ]\, (1-\gamma_5)$ in the vertex (\ref{Z2fbarfNC:Vertex}) do not contribute to the $Z\to\nu\bar\nu$ amplitude. Thus the vertex (\ref{Z2fbarfNC:Vertex}) has the same form as the SM vertex $\frac{ig}{2\cos\theta_W}\,\gamma^\mu\,(g_V-g_A\gamma_5)$ \cite{Giunti:2007ry,Novikov:1999af} with
\begin{eqnarray}
g_V &=& 1
   -\frac{1}{2}
   \exp\left(\frac{i M_Z\, p \cos\vartheta }{2\Lambda _{\rm{NC}}^2}\right)
   \nonumber\\
  & +&
   2 i \kappa \sin^2\theta_W  \sin \left(\frac{M_Z\, p \cos\vartheta }{2
   \Lambda _{\rm{NC}}^2}\right)~,
\label{Z2fbarfNC:gV} \\
g_A &=& 1 -\frac{1}{2}
\exp\left(\frac{i M_Z\, p \cos\vartheta}{2 \Lambda _{\rm{NC}}^2}\right)~.
\label{Z2fbarfNC:gA}
\end{eqnarray}
The temporary component $\vec E_\theta$ of $\theta$ is reduced from equations above since for the Z--boson at rest we have
\begin{equation}
p'\theta\, p = -M_Z\, \vec{p}\cdot {\vec{E}_{\theta}}
= - \frac{M_Z\; p \cos\vartheta}{\Lambda_{\rm{NC}}^2}~.
\label{p'tp}
\end{equation}
with $|\vec{E}_{\theta}|=1/\Lambda_{\rm NC}^2$ and $\vartheta$ the angle between $\vec p$ and $\vec E_\theta$ respectively.

Using $Z\nu\bar\nu$ vertex (\ref{Z2fbarfNC:Vertex}),  we obtain the following $Z\to \nu\bar{\nu}$ partial width \cite{Horvat:2012vn}
\begin{equation}
\begin{split}
&\Gamma(Z\to\nu\bar\nu)=
\Gamma_{\rm SM}(Z\to\nu\bar\nu)
\\&
+\frac{\alpha}{3 M_Z |\vec{E_\theta}|}
   \Bigg[\kappa  \left(1 -\kappa +\kappa\cos 2 \theta _W \right) 
   \sec ^2\theta_W\cos\left(\frac{M_Z^2 |\vec{E_\theta}|}{4}\right)
\\&
   -8 \csc ^2 2 \theta _W
 \Bigg]
 \cdot \sin\left(\frac{M_Z^2 |\vec{E_\theta}|}{4}\right)
\\&
+\frac{\alpha M_Z}{12}   \bigg[-2 \kappa ^2
+(\kappa  (2 \kappa -1)+2) \sec^2\theta _W+2 \csc^2\theta _W\bigg],
\end{split}
\label{rateZnunu}
\end{equation}
whose NC part vanishes when $\vec{E}_\theta\to 0$, i.e. for vanishing $\theta$ or space-like noncommutativity, but not light-like \cite{Gomis:2000zz,Aharony:2000gz}.

\begin{figure}[top]
\begin{center}
\includegraphics[width=8cm,angle=0]{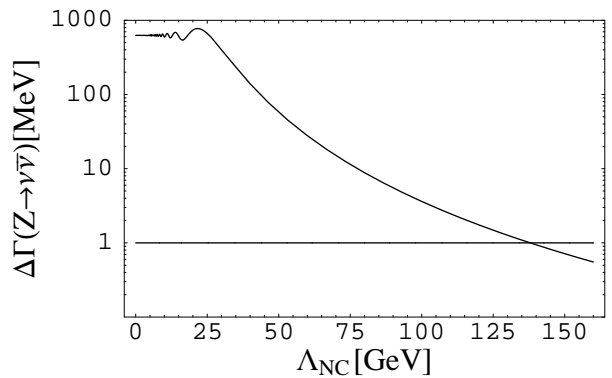}
\end{center}
\caption{$\Delta\Gamma(Z\to\nu\bar\nu)$ decay width vs. $\Lambda_{\rm{NC}}$.}
\label{fig:Z2nubarnuNC}
\end{figure}
A comparison of the experimental Z decay width $\Gamma_{\rm invisible}=(499.0\pm 1.5)$ MeV \cite{PDG2011} with its SM theoretical counterpart, allows us to set a constraint $\Gamma(Z\to\nu\bar\nu) - \Gamma_{\rm SM}(Z\to\nu\bar\nu) \lesssim 1$ MeV, from where a bound on the scale of noncommutativity
$\Lambda_{\rm{NC}} = {|\vec{E_\theta}|^{-1/2}}\stackrel{>}\sim 140$ GeV is obtained  (see Fig.~\ref{fig:Z2nubarnuNC}), for the choice $\kappa =1$.

\section{Discussion and conclusions}

We have presented the tree level cosmogenic neutrinos ($\nu$'s) scatterings: $\nu N\to\nu\,+\, anything$ and particle decays: ($(\gamma_{\rm pl},Z)\to\nu\bar\nu$) in the covariant $\theta$-exact noncommutative quantum gauge theory based on Seiberg-Witten maps and enveloping algebra formalism. 

In the energy range of interest, $10^{10}$ to ${10^{11}}$ GeV,  where there is always energy of the system $(E)$ larger than the NC scale 
$(E/\Lambda_{\rm NC}>1)$, the perturbative expansion in terms of $\Lambda_{\rm NC}$ retains no longer its meaningful character, thus it is
forcing us to resort to those NC field-theoretical frameworks involving the 
full $\theta$-resummation. Our numerical estimates of the contribution to the processes coming from the photon exchange, pins impeccably down a lower bound on $\Lambda_{\rm NC}$ to be as high as around up to ${\cal O}(10^6)$ GeV,
depending on the cosmogenic $\nu$-flux. 

For above analysis it was necessary to use results of \cite{Horvat:2011qn} which shows explicitly that the ``no-go theorem''  \cite{Chaichian:2009uw} is certainly not applicable to our SW-map based $\theta$-exact models of the NCGFT. Namely, it is known to be impossible in noncommutative geometry to directly form tensor products from the NC fields as long as there is no additional underlaying mathematical structure.  The SW-map based models do however have an additional underlying mathematical structure: They can be understood as the deformation quantization of ordinary fiber bundles over a Poisson manifold. With this additional structure, tensor products  are possible and survive the quantization procedure~\cite{Jurco:2001kp}. However, the authors in \cite{Chaichian:2009uw} failed to directly form tensor products of noncommutative fields. The proof of this failure is given in \cite{Horvat:2011qn}. 

Now we first discuss $\theta$-exact computation of the
one-loop quantum correction to the $\nu$-propagator.
We in particular evaluate  the neutrino two-point function, 
and demonstrate how quantum effects in the $\theta$-exact 
SW map approach to NCGFT's, together with a combination of Schwinger,
Feynman, and  HQET parameterization, reveal a much richer structure
yielding the one-loop quantum correction in a closed form.
 
General expression for the neutrino self-energy
(\ref{sigma1AB}) contains in (\ref{A1}) both a hard $1/\epsilon$
UV term and  the UV/IR mixing term
with a logarithmic infrared singularity $\ln |\theta p|$.
Results shows complete decoupling of the UV divergent term 
from softened UV/IR mixing term and from the finite terms as well. 
Our deformed dispersion relations at both the low and high energies 
and at the leading order
do not depend on the noncommutative scale $\Lambda_{\rm NC}$.
The low energy dispersion relation \eqref{859pm59} is, 
in principle, capable of generating a direction dependent superluminal velocity, 
this can be seen clearly from the maximal attainable velocity of the neutrinos
\begin{equation}
\frac{{\rm v}_{max}}{c}=\frac{dE}{d|\vec p|}\sim \sqrt{
1+\left(859\pm 59\right)
\sin^2\vartheta}\,,
\label{varepsilon}
\end{equation}
where $\vartheta$ is the angle with respect to the direction perpendicular 
to the NC plane. This gives one more example how such spontaneous $\theta$-background breaking of Lorentz symmetry could affect  the particle kinematics through quantum corrections, even without divergent behavior like UV/IR mixing. 
On the other hand one can also see that the magnitude of superluminosity 
is in general very large in our model as a quantum effect, thus seems contradicting various observations which suggests much smaller values 
\cite{Hirata:1987hu,Bionta:1987qt,Longo:1987ub}.   
On the other hand, note that the large superluminal velocity issue may also be reduced/removed by taking into account several considerations and/or properties:\\
(1.) Selection of a constant nonzero $\theta$ 
background in this paper is due to the computational simplicity. 
The results will, however, still hold for a NC background 
that is varying sufficiently slowly with respect to the scale of
noncommutativity. There is no physics reason to expect 
$\theta$ to be a globally constant background {\it ether}.  
In fact, if the $\theta$ background is only nonzero in 
tiny regions (NC bubbles) the effects of the modified dispersion
relation will be suppressed macroscopically.
Certainly a better understanding of possible sources of NC is needed.\\
(2.) We have considered only the purely noncommutative 
neutrino-photon coupling. However, it has been pointed out that modified 
neutrino dispersion relation could open decay channels within 
the commutative standard model framework \cite{Cohen:2011hx}. 
In our case this would further provide decay channel(s) which can 
bring superluminal neutrinos to normal ones.\\
(3.) Note that the model 1 is not the only allowed deformed model with noncommutative neutrino-photon coupling. And as we have shown for our model 2, 
there could be no modified dispersion relation(s) for deformation(s) 
other than 1, therefore it is reasonable to conjecture 
that Seiberg-Witten map freedom may also serve as one possible 
remedy to this issue.\\
(4.) Our results differs with respect to \cite{Hayakawa:1999yt}
since in our case both terms are proportional to the spacetime
noncommutativity dependent $\theta$-ratio (the scale-independent structure!) factor in (\ref{degen1}), which arise from the natural non-locality of our actions.
Besides the divergent terms, a new spinor structure $(\theta\theta p)$
with finite coefficients emerges in our computation, see (\ref{sigma1AB})-(\ref{A2}).
All these structures are proportional to $p^2$, therefore if appropriate
renormalization conditions are imposed, the commutative dispersion relation
$p^2=0$ can still hold, as a part of the full set of solutions obtained in (23).\\
(5.) Finally, we mention that our approach to UV/IR mixing should not be confused with the one  based on a theory with UV completion ($\Lambda_{\rm UV} < \infty$), where a theory becomes an effective QFT, and the UV/IR mixing manifests itself via a specific relationship between the UV and the IR cutoffs
\cite{AlvarezGaume:2003mb,Abel:2006wj}.

From the same actions (\ref{Sgauge}, \ref{Sfermion}), 
but for three different cosmological laboratories, that is from UHE cosmic ray neutrino scatterings on nuclei \cite{Horvat:2010sr}, from the BBN and from the RPAI \cite{Horvat:2009cm}, we obtain very similar, a quite strong bounds on the NC scale, of the order of $10^6$ GeV. Note in particular that all results depicted in Figs.\ref{fig:ncSM-CrossSections}-\ref{fig:LambdaVsT}, and \ref{fig:Z2nubarnuNC} show closed-convergent forms.

All above summarized properties are previously unknown 
features of $\theta$-exact NC gauge field theory. 
They appear in the model with the action presented in section~4.
The alternative action, and the corresponding $\nu$-self-energy (\ref{A9}),
has less striking features, but it does have it's own advantages due to 
the absence of a hard UV divergences, and the
absence of complicated finite terms. The structure in (\ref{A9}) is
different (it is {\it NC-scale/energy} dependent) with respect to
the NC scale-independent structure from (\ref{degen1}),
as well as to the structure arising from fermion self-energy computation in
the case of $\star$-product only unexpanded theories 
\cite{Hayakawa:1999yt,Brandt:2002if}.
However, (\ref{A9}) does posses powerful UV/IR mixing effect.
This is fortunate with regard to the use of  low-energy NCQFT
as an important window  to holography \cite{Horvat:2010km} and quantum gravity \cite{Szabo:2009tn}.

\vskip3mm Work supported by the Croatian Ministry of Science, Education and
Sport project 098-0982930-2900. I would like to thank Jiangyang You for many
valuable comments/remarks.
 
Based on the plenary session talks given at the 
5th Petrov International Symposium: "High Energy Physics, Cosmology and Gravity", BITP, Kyiv, Ukraine 2012.


\begin{thebibliography}{9}

\bibitem{Seiberg:1999vs}
N.~Seiberg and E.~Witten, 
JHEP {\bf 9909} (1999) 032,
  [\href{http://xxx.lanl.gov/abs/hep-th/9908142}{{\tt hep-th/9908142}}].
  
\bibitem{Hinchliffe:2002km}
I.~Hinchliffe, N.~Kersting, and Y.~L. Ma,
 Int. J. Mod. Phys. {\bf A19} (2004),
  179--204, [\href{http://xxx.lanl.gov/abs/hep-ph/0205040}{{\tt  hep-ph/0205040}}].

\bibitem{Trampetic:2008bk}
  J.~Trampetic,
  Fortsch.\ Phys.\  {\bf 56} (2008) 521,
  [arXiv:0802.2030 [hep-ph]].
  

\bibitem{Jackiw:2001jb}
  R.~Jackiw, S.~Y.~Pi,
  Phys.\ Rev.\ Lett.  {\bf 88} (2002) 111603,
  [arXiv:hep-th/0111122].
  
\bibitem{Madore:2000en}
J.~Madore, S.~Schraml, P.~Schupp, and J.~Wess,
  Eur. Phys. J. {\bf C16} (2000) 161, 
  [\href{http://xxx.lanl.gov/abs/hep-th/0001203}{{\tt hep-th/0001203}}].

\bibitem{Jurco:2000ja}
B.~Jurco, S.~Schraml, P.~Schupp, and J.~Wess, 
 Eur. Phys. J.  {\bf C17} (2000) 521,
  [\href{http://xxx.lanl.gov/abs/hep-th/0006246}{{\tt hep-th/0006246}}].

\bibitem{Jurco:2000fb}
  B.~Jurco and P.~Schupp,
 Eur.\ Phys.\ J.  {\bf C14} (2000) 367, [arXiv:hep-th/0001032].

\bibitem{Jurco:2001my}
B.~Jurco, P.~Schupp, and J.~Wess, 
  Nucl. Phys.  {\bf B604} (2001) 148,
  [\href{http://xxx.lanl.gov/abs/hep-th/0102129}{{\tt  hep-th/0102129}}].

\bibitem{Jurco:2001rq}
B.~Jurco, L.~Moller, S.~Schraml, P.~Schupp, and J.~Wess,
 Eur. Phys. J.  {\bf C21} (2001) 383,
  [\href{http://xxx.lanl.gov/abs/hep-th/0104153}{{\tt hep-th/0104153}}].

\bibitem{Calmet:2001na}
X.~Calmet, B.~Jurco, P.~Schupp, J.~Wess, and M.~Wohlgenannt,
   Eur. Phys. J.  {\bf C23} (2002) 363,
   [\href{http://xxx.lanl.gov/abs/hep-ph/0111115}{{\tt hep-ph/0111115}}].

\bibitem{Chaichian:2009uw}
 M.~Chaichian, P.~Presnajder, M.~M.~Sheikh-Jabbari, A.~Tureanu,
  Phys.\ Lett.\  {\bf B683}, 55-61 (2010),
  [arXiv:0907.2646 [hep-th]].
  
\bibitem{Horvat:2011qn}
R.~Horvat, A.~Ilakovac, P.~Schupp, J.~Trampeti\'{c}, and J.~You,
Phys.\ Lett.\  {\bf B715}, 340-347 (2012),
\href{http://arxiv.org/abs/1109.3085}{{\ttfamily arXiv:1109.3085}}.

\bibitem{Behr:2002wx}
W.~Behr, N.~Deshpande, G.~Duplan\v{c}i\'c, P.~Schupp, J.~Trampeti\'c, and J.~Wess, 
  Eur. Phys. J.  {\bf C29} (2003) 441,
  [\href{http://xxx.lanl.gov/abs/hep-ph/0202121}{{\tt hep-ph/0202121}}].

\bibitem{Deshpande:2001mu}
  N.~Deshpande and X.~He,
  Phys.\ Lett. {\bf B533} (2002) 116,
    [hep-ph/0112320].

\bibitem{Duplancic:2003hg}
  G.~Duplancic, P.~Schupp and J.~Trampetic,
  Eur.\ Phys.\ J. {\bf C32} (2003) 141,
  [arXiv:hep-ph/0309138].

\bibitem{Aschieri:2002mc}
P.~Aschieri, B.~Jurco, P.~Schupp, and J.~Wess,
Nucl. Phys.  {\bf B651} (2003) 45,
 [\href{http://xxx.lanl.gov/abs/hep-th/0205214}{{\tt hep-th/0205214}}.

\bibitem{Melic:2005fm}
B.~Melic, K.~Passek-Kumericki, J.~Trampetic, P.~Schupp, and M.~Wohlgenannt,
 Eur. Phys. J.  {\bf C42} (2005) 483--499,
  [\href{http://xxx.lanl.gov/abs/hep-ph/0502249}{{\tt hep-ph/0502249}}].
  
\bibitem{Melic:2005am}
B.~Melic, K.~Passek-Kumericki, J.~Trampetic, P.~Schupp, and
M.~Wohlgenannt,
Eur. Phys. J.  {\bf C42} (2005) 499--504,
  [\href{http://xxx.lanl.gov/abs/hep-ph/0503064}{{\tt hep-ph/0503064}}].
  
\bibitem{Filk:1996dm}
T.~Filk, 
Phys. Lett. {\bf B376} (1996) 53.

\bibitem{MS-R} 
C.P.~Martin, D.~Sanchez-Ruiz, 
Phys. Rev. Lett.  {\bf 83} (1999) 476--479,
 [\href{http://xxx.lanl.gov/abs/hep-th/9903077}{{\tt hep-th/9903077}}].

\bibitem{Minwalla:1999px}
  S.~Minwalla, M.~Van Raamsdonk and N.~Seiberg,
  JHEP {\bf 0002}, 020 (2000),
  [arXiv:hep-th/9912072].
  
\bibitem{Matusis:2000jf}
A.~Matusis, L.~Susskind, and N.~Toumbas, 
  JHEP {\bf 0012} (2000) 002,
  [\href{http://xxx.lanl.gov/abs/hep-th/0002075}{{\tt hep-th/0002075}}].
  %
\bibitem{Grosse:2004yu}
H.~Grosse and R.~Wulkenhaar, 
{\em Commun. Math. Phys.} {\bf 256} (2005) 305--374,
  [\href{http://xxx.lanl.gov/abs/hep-th/0401128}{{\tt hep-th/0401128}}].
  
\bibitem{Magnen:2008pd}
  J.~Magnen, V.~Rivasseau and A.~Tanasa,
  Europhys.\ Lett.\  {\bf 86} (2009) 11001,
  [arXiv:0807.4093 [hep-th]].
  
\bibitem{arXiv:1111.5553}
  S.~Meljanac, A.~Samsarov, J.~Trampetic and M.~Wohlgenannt,
  JHEP {\bf 1112} (2011) 010, arXiv:1111.5553 [hep-th].
  %
\bibitem{Vilar:2009er}
L.~C.~Q. Vilar, O.~S. Ventura, D.~G. Tedesco, and V.~E.~R. Lemes,
{\bf PoS ISFTG (2009) 071},
 \href{http://xxx.lanl.gov/abs/0902.2956}{{\tt 0902.2956}}.


\bibitem{Martin:2002nr}
C.~P. Martin, 
  Nucl. Phys.  {\bf B652} (2003) 72,
  [\href{http://xxx.lanl.gov/abs/hep-th/0211164}{{\tt hep-th/0211164}}].
  
\bibitem{Brandt:2003fx}
  F.~Brandt, C.~P.~Martin, and F.~R.~Ruiz,
  JHEP {\bf 0307}, 068 (2003),
  [arXiv:hep-th/0307292].
        
 \bibitem{Bichl:2001cq}
A.~Bichl, J.~Grimstrup, H.~Grosse, L.~Popp, M.~Schweda, and R.~Wulkenhaar,
  JHEP {\bf 0106} (2001) 013,
  [\href{http://xxx.lanl.gov/abs/hep-th/0104097}{{\tt hep-th/0104097}}].
  

\bibitem{Buric:2006wm}
M.~Buric, V.~Radovanovic, and J.~Trampetic,
  JHEP {\bf  0703} (2007) 030,
   [\href{http://xxx.lanl.gov/abs/hep-th/0609073}{{\tt hep-th/0609073}}].

\bibitem{Latas:2007eu}
D.~Latas, V.~Radovanovic, and J.~Trampetic,
  Phys. Rev. {\bf D76} (2007) 085006,
  [\href{http://xxx.lanl.gov/abs/hep-th/0703018}{{\tt hep-th/0703018}}].
  
\bibitem{Martin:2007wv}
  C.~P.~Martin, C.~Tamarit,
  Phys.\ Lett. {\bf B658}, 170 (2008),
 [arXiv:0706.4052 [hep-th]].

\bibitem{Buric:2007ix}
  M.~Buric, D.~Latas, V.~Radovanovic and J.~Trampetic,
  Phys.\ Rev. {\bf D77} (2008) 045031,
  [arXiv:0711.0887].
  %
\bibitem{Martin:2009sg}
  C. Martin, C.~Tamarit,
  Phys.\ Rev.\  {\bf D80}, 065023 (2009),
  [arXiv:0907.2464 [hep-th]].
  %
\bibitem{Martin:2009vg}
  C.~P.~Martin, C.~Tamarit,
  JHEP {\bf 0912} (2009) 042,
  [arXiv:0910.2677 [hep-th]].
    %
\bibitem{Tamarit:2009iy}
  C.~Tamarit,
  Phys.\ Rev. {\bf D81} (2010) 025006,
  [arXiv:0910.5195].
  %
\bibitem{Buric:2010wd}
  M.~Buric, D.~Latas, V.~Radovanovic and J.~Trampetic,
  Phys.\ Rev. {\bf D83} (2011) 045023,
  arXiv:1009.4603.

\bibitem{Schupp:2002up}
P.~Schupp,J.~Trampetic, J.~Wess, and G.~Raffelt,
   Eur. Phys. J. {\bf C36} (2004) 405,
     [\href{http://xxx.lanl.gov/abs/hep-ph/0212292}{{\tt hep-ph/0212292}}].

\bibitem{Minkowski:2003jg}
P.~Minkowski, P.~Schupp, and J.~Trampetic,
   Eur. Phys. J. {\bf C37}  (2004) 123,
   [\href{http://xxx.lanl.gov/abs/hep-th/0302175}{{\tt hep-th/0302175}}].
   
\bibitem{Ohl:2004tn}
T.~Ohl and J.~Reuter,
   Phys. Rev. {\bf D70} (2004) 076007,
  [\href{http://xxx.lanl.gov/abs/hep-ph/0406098}{{\tt hep-ph/0406098}}].
  
\bibitem{Alboteanu:2006hh}
A.~Alboteanu, T.~Ohl, and R.~Ruckl,
 Phys. Rev. {\bf D74} (2006) 096004, ibid {\em Acta Phys. Polon.} {\bf B38} (2007) 3647, [\href{http://xxx.lanl.gov/abs/hep-ph/0608155}{{\tt hep-ph/0608155}}].



\bibitem{Buric:2007qx}
M.~Buric, D.~Latas, V.~Radovanovic, and J.~Trampetic,
  Phys. Rev. {\bf D75} (2007) 097701,
  \href{http://xxx.lanl.gov/abs/hep-ph/0611299}{{\tt hep-ph/0611299}}.
  
\bibitem{Ohl:2010zf}
  T.~Ohl, C.~Speckner,
  Phys.\ Rev.\  {\bf D82 } (2010)  116011,
  [arXiv:1008.4710 [hep-ph]]. 
  
\bibitem{Melic:2005hb}
  B.~Melic, K.~Passek-Kumericki, and J.~Trampetic,
  Phys.\ Rev. {\bf D72}, 054004 (2005),
 [arXiv:hep-ph/0503133].
 
\bibitem{Melic:2005su}
 B.~Melic, K.~Passek-Kumericki and J.~Trampetic,
  Phys.\ Rev. {\bf D72} (2005) 057502,
  [arXiv:hep-ph/0507231].
 %
\bibitem{Tamarit:2008vy}
  C.~Tamarit and J.~Trampetic,
  Phys.\ Rev. {\bf D79}, 025020 (2009),
  [arXiv:0812.1731 [hep-th]].
  %
\bibitem{Horvat:2009cm}
  R.~Horvat, J.~Trampetic,
  Phys.\ Rev.{\bf D79} (2009) 087701,
  [arXiv:0901.4253 [hep-ph]].
 
\bibitem{Schupp:2008fs}
P.~Schupp and J.~You,
 JHEP {\bf 0808} (2008) 107,
  [\href{http://xxx.lanl.gov/abs/0807.4886}{{\tt 0807.4886}}].

\bibitem{arXiv:1109.2485}
  R.~Horvat, A.~Ilakovac, J.~Trampetic and J.~You,
  JHEP {\bf 1112} (2011) 050, arXiv:1109.2485 [hep-th].
  
\bibitem{arXiv:1111.4951}
  R.~Horvat, A.~Ilakovac, P.~Schupp, J.~Trampetic and J.~You,
 JHEP {\bf 1204} (2012) 108,
  arXiv:1111.4951 [hep-th].
  
\bibitem{Zeiner:2007}
J.~Zeiner,
\newblock PhD thesis.

\bibitem{Horvat:2010sr}
  R.~Horvat, D.~Kekez and J.~Trampetic,
  Phys.\ Rev. {\bf D83} (2011) 065013,
 arXiv:1005.3209 [hep-ph].
  %
\bibitem{Horvat:2011iv}
  R.~Horvat, D.~Kekez, P.~Schupp, J.~Trampetic, J.~You,
 Phys. Rev. {\bf D84} (2011) 045004,
[\href{http://xxx.lanl.gov/abs/1103.3383}{arXiv:1103.3383 [hep-ph]}.

\bibitem{Horvat:2011wh}
  R.~Horvat, J.~Trampetic,
 Phys. Lett. {\bf B710} (2012) 219,
  arXiv:1111.6436 [hep-ph].



\bibitem{Okawa:2001mv}
Y.~Okawa and H.~Ooguri,
 Phys. Rev. {\bf D64} (2001) 046009,
  [\href{http://xxx.lanl.gov/abs/hep-th/0104036}{{\tt hep-th/0104036}}].
 
\bibitem{Martin:2012aw} 
  C.~P.~Martin,
  arXiv:1206.2814 [hep-th].
  %
\bibitem{Horvat:2010km}
  R.~Horvat and J.~Trampetic,
  JHEP {\bf 1101} (2011) 112,
  [arXiv:1009.2933 [hep-ph]].
\bibitem{Gomis:2000zz} 
  J.~Gomis and T.~Mehen,
  Nucl.\ Phys. {\bf B591}, 265 (2000),
  [hep-th/0005129].
  
\bibitem{Aharony:2000gz} 
  O.~Aharony, J.~Gomis and T.~Mehen,
  JHEP {\bf 0009}, 023 (2000),
  [hep-th/0006236].
  
\bibitem{Aglietta:2007yx}
  J.~Abraham {\it et al.}  [Pierre Auger Collaboration],
  Astropart.\ Phys.\  {\bf 29}, 243 (2008), [0712.1147 [astro-ph]].
  
\bibitem{Fodor:2003ph}
  Z.~Fodor, S.~D.~Katz, A.~Ringwald and H.~Tu,
  JCAP {\bf 0311}, 015 (2003).
  [arXiv:hep-ph/0309171].
%
\bibitem{Protheroe:1995ft}
  R.~J.~Protheroe and P.~A.~Johnson,
  Astropart.\ Phys.\  {\bf 4}, 253 (1996),  [arXiv:astro-ph/9506119].
  
\bibitem{Kravchenko:2002mm}
  I.~Kravchenko {\it et al.},
  Astropart.\ Phys.\  {\bf 20}, 195 (2003),
  [arXiv:astro-ph/0206371].
  
  \bibitem{Hayakawa:1999yt}
M.~Hayakawa,
  Phys.\ Lett. {\bf B478}, 394 (2000), [arXiv:hep-th/9912094].
  
\bibitem{Chaichian:2001mu}
  M.~Chaichian, P.~Presnajder, M.~M.~Sheikh-Jabbari, A.~Tureanu,
  Phys.\ Lett.\  {\bf B526}, 132-136 (2002),
  [hep-th/0107037].
  
\bibitem{Jurco:2001kp}
  B.~Jurco, P.~Schupp, J.~Wess,
  Lett.\ Math.\ Phys.\  {\bf 61}, 171-186 (2002),  [hep-th/0106110].
  
\bibitem[Bordemann et al.(2007)]{2007arXiv0711.2965B} 
M.~Bordemann, N.~Neumaier, S.~Waldmann, and S.~Weiss, arXiv:0711.2965.

\bibitem[Bursztyn et al.(2009)]{2009arXiv0909.4259B} 
H.~Bursztyn, V.~Dolgushev, and S.~Waldmann, arXiv:0909.4259.

\bibitem{Schupp:2001we}
P.~Schupp,
PrHEP-hep2001/238 (2001),
[\href{http://xxx.lanl.gov/abs/hep-th/0111038}{hep-th/0111038}].

\bibitem{Grozin:2000cm}
A.~G. Grozin, 
  \href{http://xxx.lanl.gov/abs/hep-ph/0008300}{{\tt hep-ph/0008300}}.
       %
  %


   %
\bibitem{Giunti:2007ry}
C.~Giunti and C.~W. Kim, 
{\it Fundamentals of Neutrino Physics and Astrophysics}, Oxford, UK:U.Pr.(2007)710 p.

\bibitem{Novikov:1999af}
V.~A. Novikov, L.~B. Okun, A.~N. Rozanov, and M.~I. Vysotsky,
 \href{http://dx.doi.org/10.1088/0034-4885/62/9/201}{
  Rept. Prog. Phys. {\bfseries 62} (1999) 1275--1332},
\href{http://arxiv.org/abs/hep-ph/9906465}{{\ttfamily arXiv:hep-ph/9906465}}.

\bibitem{Horvat:2012vn} 
  R.~Horvat, A.~Ilakovac, D.~Kekez, J.~Trampetic and J.~You,
  arXiv:1204.6201 [hep-ph].

\bibitem{PDG2011}
K. Nakamura et al. (Particle Data Group), Journal of Physics  {\bf G37}, 075021 (2010) and 2011 partial update for the 2012 edition.
    %
  

    
 
\bibitem{Hirata:1987hu}
  K.~Hirata {\it et al.}  [KAMIOKANDE-II Collaboration],
  Phys.\ Rev.\ Lett.\  {\bf 58} (1987) 1490.

\bibitem{Bionta:1987qt}
  R.~M.~Bionta, G.~Blewitt, C.~B.~Bratton, D.~Casper, A.~Ciocio, R.~Claus, B.~Cortez,  M.~Crouch, S.T. Dye, S. Errede {\it et al.},
  Phys.\ Rev.\ Lett.\  {\bf 58} (1987) 1494.

\bibitem{Longo:1987ub}
  M.~J.~Longo,
  Phys.\ Rev. {\bf D36} (1987) 3276.
  
\bibitem{Cohen:2011hx}
  A.~G.~Cohen and S.~L.~Glashow,
  Phys.\ Rev.\ Lett.\  {\bf 107} (2011) 181803,
  [arXiv:1109.6562 [hep-ph]].
  
\bibitem{AlvarezGaume:2003mb}
  L.~Alvarez-Gaume and M.~A.~Vazquez-Mozo,
  Nucl.\ Phys. {\bf B668}, 293 (2003),
  [arXiv:hep-th/0305093].
  %
\bibitem{Abel:2006wj}
  S.~A.~Abel, J.~Jaeckel, V.~V.~Khoze and A.~Ringwald,
  JHEP {\bf 0609}, 074 (2006),
  [arXiv:hep-ph/0607188].
     
%
  
\bibitem{Brandt:2002if}
  F.~T.~Brandt, A.~K.~Das, J.~Frenkel,
  Phys.\ Rev.\  {\bf D66 } (2002)  065017, [hep-th/0206058]. 
  %
\bibitem{Szabo:2009tn}
  R.~J.~Szabo,
  Gen.\ Rel.\ Grav.\  {\bf 42 } (2010)  1-29,  [arXiv:0906.2913 [hep-th]].
  

  
\end{thebibliography}
\end{document}